\documentclass[11pt,twoside]{article}
\usepackage{asp2004}
\usepackage{psfig}
\usepackage{epsf}
\usepackage{graphics}
\usepackage{lscape}
\def\edcomment#1{\iffalse\marginpar{\raggedright\sl#1\/}\else\relax\fi}
\reversemarginpar

\begin{document}
\title{The role of the SW Sextantis stars in the picture of CV evolution}
\author{P. Rodr\'\i guez-Gil}
\affil{Department of Physics, University of Warwick, Coventry CV4 7AL, UK}

\begin{abstract}
The SW Sextantis stars are nova-like cataclysmic variables (CVs) which exhibit intricate behaviour which is still a matter of debate. For years the common belief has been that these systems were the``outlaws'' of the CVs, mainly because of the lack of an overall understanding. We are now realising that a large percentage ($\sim 30$\%) of all nova-likes in the $\sim 3-4$ hour period range are SW Sex stars. The study of the dominance of this class just above the period gap will provide clues on the evolution of CVs. Here we present the discovery of 7 new SW Sex stars from the CVs found in the Hamburg Quasar Survey.
\end{abstract}
\thispagestyle{plain}

%_______________________________________________________________________________

\section{CV evolution: theory {\it vs} observations}

The currently accepted theory of CV evolution relies on the disrupted magnetic braking model. For $P_\mathrm{orb} > 3$ h angular momentum loss by stellar magnetic braking dominates, whereas emission of gravitational radiation is believed to be the only mechanism at $P_\mathrm{orb} < 3$ h. %\citep{rappaportetal83-1,spruit+ritter83-1,paczynski+sienkiewicz83-1}.
%The evolution of CVs can be driven by two angular momentum loss mechanisms. Stellar magnetic braking dominates in CVs whose Roche-lobe filling donor stars still have a radiative core ($P_\mathrm{orb} > 3$ h). Once the donor stars become fully convective at $P_\mathrm{orb} \sim 3$ h, magnetic braking ceases. For   loss mechanism, resulting in longer evolution time scales.
The observational input does not confirm a number of predictions of this theory, like the minimum period a CV can have (65 min; we observe $\sim 76$ min) and a pile-up of systems close to this minimum. To make things even worse, there is no observational evidence for a discontinous change in the spin-down rate due to magnetic braking between late-type field stars that are fully convective and those that have a radiative core \citep{andronovetal03-1}.

%_______________________________________________________________________________

\section{The Hamburg Quasar Survey sample}

The discrepancies between theory and observation are mainly due to strong selection effects. It is then fundamental to obtain an unbiased CV sample to compare its orbital period distribution with that predicted by theory, which will be a fundamental test. To do so, we have started a large scale search for new CVs in the Hamburg Quasar Survey \citep[HQS;][]{hagenetal95-1}, where CVs were selected according to their spectroscopic fingerprint: broad emission or absorption lines, or a combination of the two.

%_______________________________________________________________________________

\section{The impact of SW Sex stars in the HQS sample}

\begin{figure}[!h]
\plotfiddle{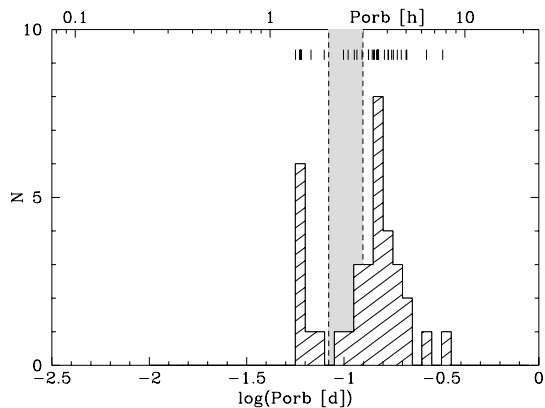}{5cm}{-90}{27}{27}{-110}{150}
\caption{\label{fig1} Period distribution of the new HQS CVs.}
\end{figure}

\begin{table}
\caption{\label{Table1}New SW Sex stars from the HQS}
\smallskip
\begin{center}
{\small
\begin{tabular}{lcc}
\tableline
\noalign{\smallskip}
Object~~~~~~~~~~~ & $P_\mathrm{orb}$(h)~~~~~~~~~~~ & Eclipsing\\   
\noalign{\smallskip}
\tableline
\noalign{\smallskip}
HS\,0129+2933~~~~~~~~~~~   & 3.35~~~~~~~~~~~ & Yes \\
HS\,0220+0603~~~~~~~~~~~   & 3.58~~~~~~~~~~~ & Yes \\
HS\,0357+0614~~~~~~~~~~~   & 3.59~~~~~~~~~~~ & No  \\
HS\,0455+8315~~~~~~~~~~~   & 3.57~~~~~~~~~~~ & Yes \\
HS\,0551+7241~~~~~~~~~~~   & 3.42~~~~~~~~~~~ & No  \\
HS\,0728+6738~~~~~~~~~~~   & 3.21~~~~~~~~~~~ & Yes \\
HS\,1813+6122~~~~~~~~~~~   & 3.55~~~~~~~~~~~ & No  \\
\noalign{\smallskip}
\tableline
\end{tabular}
}
\end{center}
\end{table}
About 20\% of the 35 new HQS CVs for which we have measured $P_\mathrm{orb}$ are SW Sex stars (Table~\ref{Table1}). Striking is the concentration of these CVs within a narrow period range between 3 and 4 hours. If we take into account all the SW Sex stars known to date, they represent $\sim 30$\% of all nova-likes in this period range. The period distribution of the 35 new CVs for which we have provided an accurate orbital period is shown in Fig.~\ref{fig1}. The ``SW Sex spike'' is evident.  
It has been suggested \citep{rodriguez-giletal01-1,rodriguez-gil03-1} that moderately strong magnetic fields might play a fundamental role in the evolution of CVs around the upper edge of the period gap. The confirmation of a probable magnetic nature will radically change the overall picture of the CV population, since in such a case the majority of CVs within $\sim 3-4$ h will be magnetic accretors, in stark contrast with the $\sim 3$\% of isolated magnetic white dwarfs. This will be very important for the current theory of CV evolution, as it will be clear indication magnetic CVs evolve differently from non-magnetic ones.

%In few years the SW Sextantis stars have changed their role of "outsider" CVs to an important position in the CV evolution scenario. Much effort towards a complete understanding of the elusive SW Sex phenomenon is therefore necessary.

\end{document}